\documentclass[preprint,prl,showpacs,preprintnumbers,amsmath,amssymb,floatfix]{revtex4}

\usepackage{graphicx}\usepackage{dcolumn}
\usepackage{bm}

\begin{document}

\preprint{APS/123-QED}

\title{A vapor cell based on dispensers for laser spectroscopy}% Force line breaks with \\

\author{E. M. Bridge}
\author{J. Millen}
\author{C. S. Adams}
\author{M. P. A. Jones}
 \affiliation{Department of Physics, Durham University, Rochester Building, South Road, Durham, DH1 3LE, U.K.}

\newcommand{\probe}{$5{\rm s}^{2}\,^1{\rm S}_0 \rightarrow 5{\rm s}5{\rm p}\,^1{\rm P}_1$~}

\date{\today}% It is always \today, today,

\begin{abstract}
We describe a simple strontium vapor cell for laser spectroscopy experiments. Strontium vapor is produced using an electrically heated commercial dispenser source. The sealed cell operates at room temperature, and without a buffer gas or vacuum pump.  The cell was characterised using laser spectroscopy, and was found to offer stable and robust operation, with an estimated lifetime of $>10,000$ hours. By changing the dispenser, this technique can be readily extended to other alkali and alkaline
earth elements. \end{abstract}

\maketitle

\section{Introduction}
The laser spectroscopy of thermal vapors is central to atomic physics research, and is also an essential tool for laser stabilization in areas such as laser cooling and trapping. Most  experiments are carried out using the heavier alkali metals (Rb and Cs), which have sufficient vapor pressure at room temperature that a simple glass cell can be used. However, for some experiments such as high-precision frequency metrology \cite{ludlow08,blatt08} and studies of ultra-cold plasmas \cite{simien04}, other elements such as strontium offer unique advantages.  Making vapor cells for other  elements  is complicated not only by the need to reach higher temperatures, but also because the hot vapor attacks commonly used materials such as glass, quartz and  copper.  As a result,  these cells are typically more complex, employing buffer gases, water cooling and  sapphire windows \cite{gallagher95,philip07,li04}. Here we describe  a simple and compact strontium spectroscopy cell based on commercially produced dispenser sources. The cell operates at room temperature,  and no vacuum pump is needed during normal operation, making the cell simpler to construct and operate than a buffer gas cell or atomic beam apparatus. The design could be adapted to work with other elements simply by changing the dispenser \cite{dispensernote}. 

\section{Cell design and construction}
\begin{figure}
\centering
\includegraphics{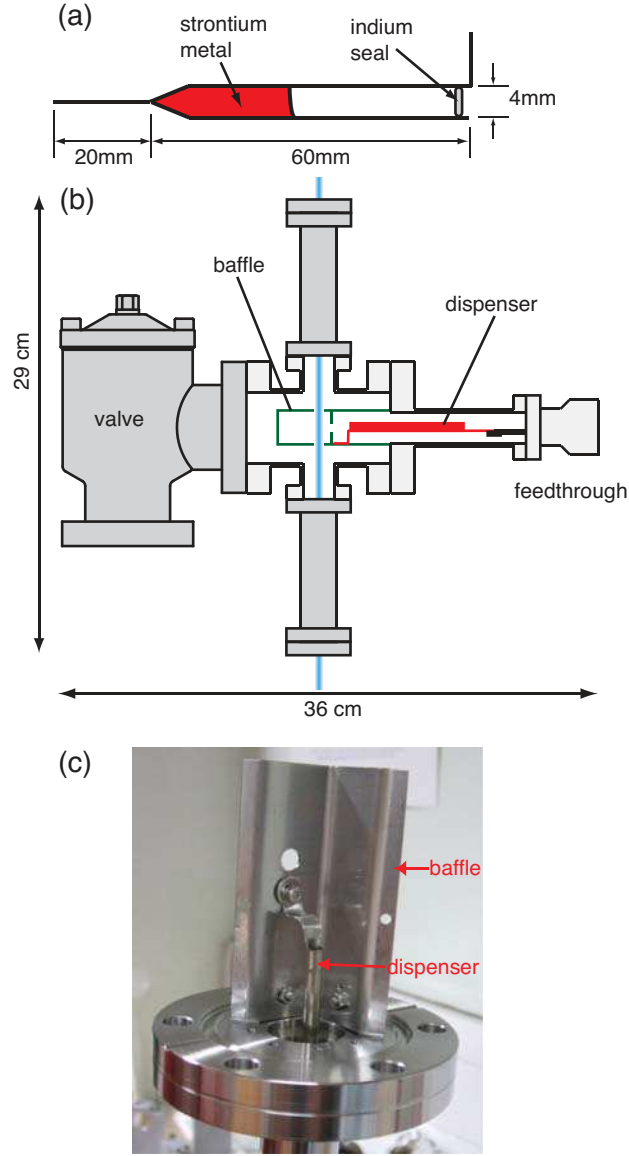}
\caption{(Color online) (a) Cross-section of the dispenser. (b) Partial sectional drawing of the cell, showing the internal layout. The cell is based on a standard DN40CF/DN16CF reducing cross. (c) The dispenser with the cell body and part of the baffle removed.\label{fig1}}
\end{figure}
The dispenser source and the design of the cell are shown in Fig.~\ref{fig1}. The strontium dispenser (Alvatec AS-Sr-500-F)  consists of a small stainless steel tube filled with 500\,mg of strontium metal under an argon atmosphere \cite{dispensernote}.   The tube is crimped closed at one end, and the other end is sealed by an indium plug. Flattened tags are provided at each end for mounting and electrical connection. With the indium seal intact, the dispensers can be easily handled under ambient conditions. Once mounted in the cell and under vacuum, the dispenser is heated by passing an electrical current through the steel tube. During a preliminary activation step, the dispenser is heated until the indium seal melts. After activation, passing a larger current through the dispenser causes the strontium metal to evaporate and a broad, weakly collimated atomic jet is emitted from the unsealed end. 

The cell is  arranged for laser spectroscopy transverse to the jet as shown in Fig.~\ref{fig1}. The cell is designed to minimize the amount of strontium vapor reaching the windows,  and thus avoid the extra cost and complexity of sapphire windows. Optical access is provided by two standard Kodial glass DN16CF viewports. The strontium emission from the dispenser is directional, and a simple stainless steel baffle ensures that there is no direct line of sight from the dispenser to the viewports, which are mounted $\sim 10$\,cm away. The baffle is closed at the opposite end to the dispenser by a lid. This room temperature metal surface efficiently adsorbs the strontium vapor, reducing its diffusion to the cell windows. After several months of operation, no change in the viewports has been observed. The strontium metal deposited on the baffle also acts as a getter, maintaining the vacuum in the cell with no additional pumps. This gives an important reduction in the size and cost of the cell. The cell body is constructed entirely from standard Conflat-type vacuum fittings. In order to reduce the leak rate to a minimum, copper gaskets were used throughout and an all-metal valve was used to seal the cell.  
	The tabs at each end of the dispenser were used for electrical connections and to mount the dispenser inside the cell. A small hole was punched in each tab. At one end this was used to bolt the dispenser to the baffle which provided a return path for the current via the cell wall. The other end of the dispenser was attached to two pins of a UHV electrical feedthrough, allowing currents of up to 20\,A to be used. 
	
After assembly, the cell was leak tested, and then evacuated using a turbomolecular pump. After baking at  150$^\circ$C for 24 hours, the cell reached an ultimate pressure of $\sim 10^{-8}$\,mbar. Previous designs for vapor cells  frequently use a buffer gas to reduce the diffusion of strontium towards the windows. Here, no buffer gas is required removing any pressure broadening from buffer gas collisions. All the materials used in the cell are compatible with operation in  UHV conditions, and with a longer bake-out  pressures $<10^{-10}$\,mbar could be achieved. 

Before the cell is isolated from the pumps, the dispenser must be heated sufficiently for the indium seal to melt. After running 5\,A through the dispenser as specified on the manufacturer's datasheet,  the current was slowly increased. At 8.0\,A, the pressure in the cell suddenly increased from $7 \times 10^{-7}$\,mbar to $1.2 \times 10^{-5}$\,mbar as the indium seal melted and the trapped argon in the dispenser was released. The dispenser was then degassed at this current overnight until the pressure returned to $\sim 10^{-7}$\,mbar.

\section{Characterisation }

In order to characterise the performance of the cell, we performed laser absorption spectroscopy on the \probe transition at 460.7\,nm.  The 460.7\,nm light is produced by  a commercial frequency-doubled diode laser system (Toptica SHG). We measured the absorption of a weak probe beam directed through the cell as the current through the dispenser was increased.  The probe beam was linearly polarized  with  $1/e^2$ radii of $1.55\pm0.05$\,mm and $2.2\pm0.1$\,mm in the horizontal and vertical directions respectively, and a power of $187\,\mu$W. The probe beam frequency was scanned over approximately 1\,GHz around the resonance by applying a linear voltage ramp to a piezo in the master laser cavity. 

\begin{figure}
\centering
\includegraphics{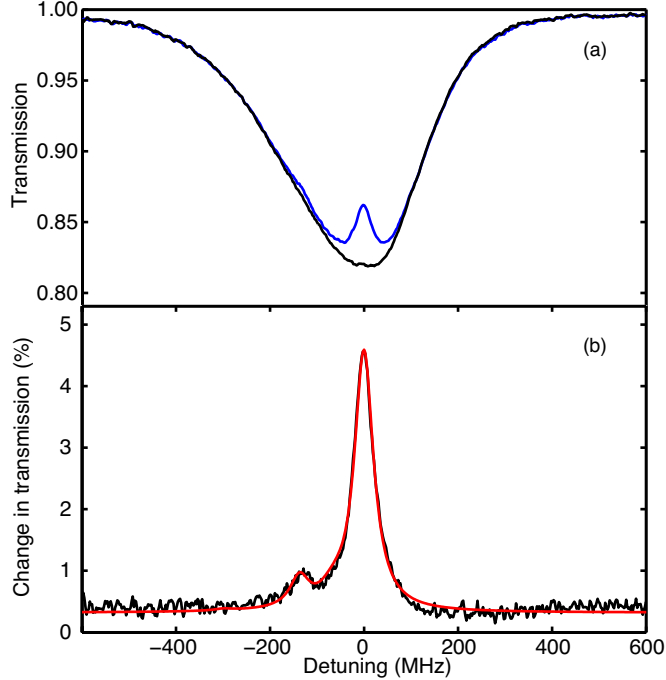}
\caption{\label{fig2} (Color online) (a) Change in probe transmission as the laser is scanned with (blue) and without (black) the pump beam. (b) Saturated absorption spectrum of the \probe transition (black), and the fit used to calibrate the frequency axis (red). The dispenser current was 15.4\,A.}
\end{figure}
The frequency axis of the scan was calibrated using saturated absorption spectroscopy. A counter-propagating pump beam was overlapped with the probe beam. The pump and probe beams had orthogonal linear polarizations.  The effect of the pump beam on the probe absorption is shown in Fig.~\ref{fig2}a, where a saturated absorption peak is clearly visible. The angle of the pump and probe beams through the cell has been adjusted to center the saturated absorption feature within the Doppler background. The difference between the probe beam absorption with and without the pump beam yields the Doppler-free spectrum shown in Fig.~\ref{fig2}b. We fit this spectrum with the sum of six Lorentzians, one for each isotope and hyperfine component and use the known natural abundances, isotope shifts and hyperfine splittings \cite{mauger08}. The remaining free parameters are an overall amplitude, a center frequency offset and the time/frequency scaling factor. With these last two fit parameters we can calibrate the frequency axis with an uncertainty of $\pm 1\%$. 

An example of the Doppler broadened absorption profile is shown in Fig.~\ref{fig2}a. The full-width half-maximum (FWHM) is $343\pm10$\,MHz, corresponding to a FWHM of $158\pm5$\,m\,s$^{-1}$ for the transverse velocity distribution. The Doppler-broadened lineshape shows a slight asymmetry that is independent of the angle of the beams through the cell, and the parameters of the laser scan. This is due to the flux from the dispenser being partially obscured by the internal baffle, and could easily be corrected with improved internal alignment. Numerical simulations of the atomic jet accurately reproduce this asymmetric lineshape with a displacement of $\sim 2$ mm between the dispenser and the baffle.

\begin{figure}
\centering
\includegraphics{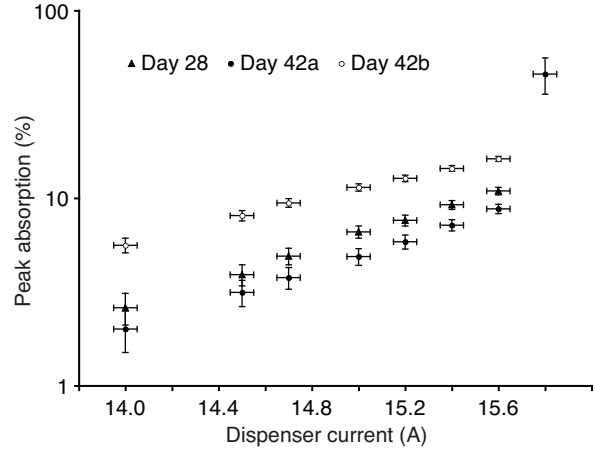}
\caption{\label{fig3}Peak absorption versus current through the dispenser. In datatset Day 42a, the dispenser current reaches the unstable region above 15.6\,A and there is a sharp increase in the absorption. On returning to the stable region (Day 42b), the gradual decline in absorption between day 28 and day 42 has been reversed.}\end{figure}

We first observe a Doppler-broadened absorption dip at a current of approximately 14.0A. The behaviour of the peak absorption as a function of the current through the dispenser is shown in Fig.~\ref{fig3}. Below 15.6A, the peak absorption is very stable and at the 0.5\% sensitivity of our measurements we observed no short-term fluctuations. However, after running the cell for approximately one month, with the dispenser switched on and off each day, a gradual decline in the peak absorption was observed. This can be reversed by increasing the current beyond 15.6\,A. Here, the strontium emission rapidly becomes unstable. If we increase the current from 15.6\,A to 15.8\,A, we initially observe a gradual increase in the peak absorption. After approximately five minutes at this current, we observed a sudden jump in the absorption from 10\% to 50\%. The peak absorption became very unstable, changing by up to 10\% on a time scale of a few seconds. We attribute this behaviour to the dispenser reaching the melting point of the strontium metal. On reducing the current below 15.6\,A, the emission stabilises again but the amount of absorption at a given current has increased as shown in Fig.~\ref{fig3}. A likely explanation for this behaviour is that in normal operation, the dispenser gradually becomes blocked by strontium metal that gradually accumulates at the ends where the dispenser is coldest. Heating the dispenser into the unstable region clears this blockage and restores the flux. 

Using the measured peak absorption and Doppler width, we can estimate the total flux and the lifetime of the dispenser. To estimate the mean longitudinal velocity of the atoms we assume that the temperature of the vapor in the dispenser is equal to the melting point of strontium (1024\,K) and that the longitudinal velocity distribution is that of an effusive atomic beam. This gives a mean longitudinal velocity of 590\,m\,s$^{-1}$. With a peak absorption of 16\,\% at 15.6\,A, we obtain a total flux of $\sim 8\times10^{13}$\,s$^{-1}$, and a lifetime of the dispenser under these conditions of $\sim 12,000$ hours, corresponding to over a year of continuous operation before the dispenser needs to be changed. So far we have successfully operated the cell for several months. This long lifetime makes the cell well suited for a wide range of experiments using thermal strontium atoms, including as a frequency reference for laser stabilisation.

\begin{acknowledgments}
The authors wish to thank Ifan Hughes for useful discussions and a careful reading of the manuscript.
\end{acknowledgments}

\end{document}